\pdfoutput=1

\documentclass[11pt]{article}

\usepackage[final]{acl}

\usepackage{times}
\usepackage{latexsym}

\usepackage[T1]{fontenc}

\usepackage[utf8]{inputenc}

\usepackage{microtype}

\usepackage{inconsolata}

\usepackage{graphicx}

\usepackage{trajan}
\usepackage{xspace}
\usepackage{booktabs}
\usepackage{graphicx}
\usepackage{wrapfig}
\usepackage{tabularx}
\usepackage{subcaption}
\usepackage{pdfpages}
\usepackage{listings}
\usepackage{amsmath}
\usepackage{multirow}
\usepackage{fontawesome5}
\usepackage{xcolor}
\usepackage{colortbl}
\usepackage{booktabs}
\usepackage{adjustbox}
\newcommand{\demoname}{{\trjnfamily SOTOPIA-S}$^4$\xspace}
\newcommand{\apiname}{{\trjnfamily SOTOPIA-API}\xspace}
\newcommand{\sotopia}{{\trjnfamily SOTOPIA}\xspace}

\definecolor{violet-5}{RGB}{132, 94, 247}

\newcommand{\aspace}{\hspace{2em}}

\newcommand{\cmu}{$^\heartsuit$}
\newcommand{\stanford}{$^\diamondsuit$}

\newcommand{\cuhk}{$^\spadesuit$}
\newcommand{\aptima}{$^\clubsuit$}

\title{\demoname: a user-friendly system for flexible, customizable, and large-scale social simulation}

\author{
Xuhui Zhou\cmu\thanks{Equal contributors.}\aspace Zhe Su\cmu\footnotemark[1]\aspace Sophie Feng\cmu \aspace Jiaxu Zhou\cuhk \aspace\\[5pt]
\textbf{~Jen-tse Huang\cuhk \aspace Hsien-Te Kao\aptima \aspace Spencer Lynch\aptima \aspace Svitlana Volkova\aptima}\\[5pt]
\textbf{~Tongshuang Sherry Wu\cmu \aspace Anita Woolley\cmu \aspace Hao Zhu\stanford \aspace Maarten Sap\cmu}\\
\vspace{4pt}
\small{\cmu Carnegie Mellon University \; \aptima Aptima \; \stanford Stanford University \; \cuhk The Chinese University of Hong Kong}
}

\begin{document}
\maketitle

\begin{abstract}
Social simulation through large language model (LLM) agents is a promising approach to explore and validate hypotheses related to social science questions and LLM agents behavior.
We present \demoname, a fast, flexible, and scalable social simulation system that addresses the technical barriers of current frameworks while enabling practitioners to generate multi-turn and multi-party LLM-based interactions with customizable evaluation metrics for hypothesis testing.
\demoname comes as a pip package that contains a simulation engine, an API server with flexible RESTful APIs for simulation management, and a web interface that enables both technical and non-technical users to design, run, and analyze simulations without programming.
We demonstrate the usefulness of \demoname with two use cases involving dyadic hiring negotiation and multi-party planning scenarios.
\end{abstract}

\section{Introduction}
\label{sec:introduction}

Social simulation has emerged as a powerful tool for understanding human behavior and social dynamics \citep{Ziems2023-llm, park2024generativeagentsimulations1000, manning2024automatedsocialsciencelanguage, gao2023s3socialnetworksimulationlarge}.
With the advancement of role-playing abilities of large language models (LLMs), we can now simulate realistic social interactions at scale \citep{zhou2024sotopia, li2023metaagents, pang2024selfalignmentlargelanguagemodels, yang2024oasisopenagentsocial}. 
However, existing frameworks require significant technical expertise to run and evaluate large-scale simulations efficiently \citep{zhou2024sotopia,park2023generativeagents, wu2023autogenenablingnextgenllm}.

We present \demoname (Simple Social Simulation System), a system designed that enables practitioners without extensive technical backgrounds to:
(1) design social simulations through \textbf{natural language specifications}, eliminating the need for programming expertise, 
(2) run multiple social interactions efficiently via automated parallelization, 
(3) customize evaluation metrics through simple configuration settings, 
and (4) manage simulated interactions and results through a web interface.

\demoname's architecture separates core simulation logic from the user interface, allowing practitioners to focus on experimental design rather than implementation details.
Specifically, we offer \apiname, a fastAPI-based protocol for simulation management.
Users can retrieve and upload characters, scenarios, evaluation metrics, and start scaled simulations through the API.
Besides the API, we also offer a web-based application for visualizing and editing scenarios, characters, and simulation results.
On the backend, the simulation engine handles complex technical aspects like asynchronous execution, LLM API management, and data persistence automatically, abstracting away the underlying complexities from the users.

To showcase the flexibility and usability of \demoname, we demonstrate two use cases.
First, we use \demoname to examine the effects of user personality in a hiring negotiation setting, by simulating multiple multi-turn dyadic interactions and evaluating the interaction outcomes.
Extending beyond dyadic interactions, we also show that \demoname can be used to simulate multi-party scenarios, where agents can act simultaneously and make contingent offers to each other.
Furthermore, we stress test the system with a large group of agents to showcase its scalability.

We release the code and a user guide at \url{https://github.com/sotopia-lab/sotopia}, a website with documentation and examples at \url{https://demo.sotopia.world}, and a video demo at \url{https://youtu.be/dZq9tNqerks}.

\section{Related Work}
\label{sec:related_work}
\definecolor{ai2_orange}{RGB}{255,150,0}
\definecolor{ai2_blue}{RGB}{0,213,255}
\definecolor{ai2_purple}{RGB}{208,191,255}
\definecolor{ai2_green}{RGB}{0,128,0}
\definecolor{ai2_red}{RGB}{139,0,0}

\begin{table*}[t]
    \centering
    \footnotesize
    \begin{tabular}{>{\centering\arraybackslash}p{2.5cm} >{\centering\arraybackslash}p{1.5cm} >{\centering\arraybackslash}p{1.5cm} >{\centering\arraybackslash}p{1.5cm} >
{\centering\arraybackslash}p{1.5cm} >
{\centering\arraybackslash}p{4.5cm}}
    \rowcolor{ai2_purple!40}\rule{0pt}{3ex}\textbf{Framework} & \textbf{NL Spec.} & \textbf{Mul-Party} & \textbf{Auto Eval} & \textbf{Web-UI} & \textbf{Social Data}\\ 
    \midrule
    OASIS \citep{yang2024oasisopenagentsocial} & \faTimes & \faCheck  & \faTimes  & \faTimes & \footnotesize Rich social scenarios and characters with relationships\\ 
    \midrule
    CrewAI \citep{crewai_website} & \faTimes & \faCheck  & \faTimes  & \faTimes & \footnotesize No existing characters and scenarios, or schema\\ 
    \midrule
    Generative Agent \citep{park2023generativeagents} & \faTimes & \faCheck  & \faTimes & \faCheck & \footnotesize Limited scenarios and characters based on the virtual town\\ 
    \midrule
    S3 \citep{gao2023s3socialnetworksimulationlarge} & \faTimes & \faCheck  & \faTimes  & \faTimes & \footnotesize Rich social scenarios and characters with relationships\\ 
    \midrule
    AutoGen \citep{wu2023autogenenablingnextgenllm} & \faTimes & \faCheck  & \faTimes  & \faTimes & \footnotesize No existing characters and scenarios, or schema\\ 
    \midrule
    \sotopia & \faTimes & \faTimes & \faCheck & \faTimes & \footnotesize Rich social scenarios and characters with relationships\\ 
    \demoname (\textbf{Ours}) & \faCheck & \faCheck & \faCheck & \faCheck & \footnotesize Rich social scenarios and characters with relationships\\ 
    \bottomrule
    \end{tabular}
    \caption{\footnotesize Comparison of multi-agent frameworks versus \demoname. NL Spec. (natural language specifications) indicates whether one can configure simulations using natural language descriptions without programming. Mul-Party (multi-party) shows if the framework supports more-than-two parties in the simulation. Auto Eval (automated evaluation) indicates built-in automated evaluation capabilities. Social Data describes the type of social interaction data provided to support the simulation.
}
\label{tab:framework_comparison}
\vspace{-10pt}
\end{table*}
\demoname takes inspiration from a long history of agent-based simulation in social sciences (\S\ref{ssec:rel-work-social-simulations}), yet differentiates itself from many existing agent simulation frameworks (\S\ref{ssec:rel-work-frameworks}).

\subsection{Social Simulation and its Applications}\label{ssec:rel-work-social-simulations}
Social simulation has been widely used to study human behavior and social phenomena. Early works focus on using rule-based agents to study social dynamics \citep{epstein1996growing, Gilbert2005AgentbasedSS}, while recent works leverage LLMs to create more realistic and complex social interactions \citep{vezhnevets2023generative, zhou2024sotopia, wang2024sotopiapiinteractivelearningsocially}. 
These simulations have been applied to various domains, including studying social norms \citep{horiguchi2024evolutionsocialnormsllm}, cultural evolution \citep{kwok2024evaluatingculturaladaptabilitylarge}, and negotiation behavior \citep{bianchi2024llmsnegotiatenegotiationarenaplatform}.
\demoname takes inspiration from these works by providing a scalable platform that enables researchers to easily design, run, and evaluate social simulations for their specific research questions.

\subsection{Multi-agent Frameworks}\label{ssec:rel-work-frameworks}
With the rise of LLMs, there has been a large increase in multi-agent frameworks that enable interactions between AI agents (Table~\ref{tab:framework_comparison}).
While frameworks like OASIS \citep{yang2024oasisopenagentsocial}, S3 \citep{gao2023s3socialnetworksimulationlarge}, and \sotopia \citep{zhou2024sotopia} provide rich social scenarios and character relationships, they lack key features like natural language configuration or web-based user interface, making it difficult for users with less programming experience to design simulations.
Another line of multi-agent frameworks, including AutoGen \citep{wu2023autogenenablingnextgenllm} and CrewAI \citep{crewai_website}, primarily focuses on problem-solving rather than social interactions. They lack pre-built social scenarios, character profiles, and relationship schemas that are essential for studying human-like social behavior.
The Generative Agents framework \citep{park2023generativeagents} includes a web interface and multi-party support but is limited by its virtual town setting. 
As shown in Table~\ref{tab:framework_comparison}, \demoname is unique in supporting natural language specifications for simulations, multi-party interactions, automated evaluation capabilities, and a web-based user interface.

\begin{figure*}[t]
    \centering
    \includegraphics[width=\textwidth, trim = 0.5cm 0cm 0.5cm 0cm, clip]{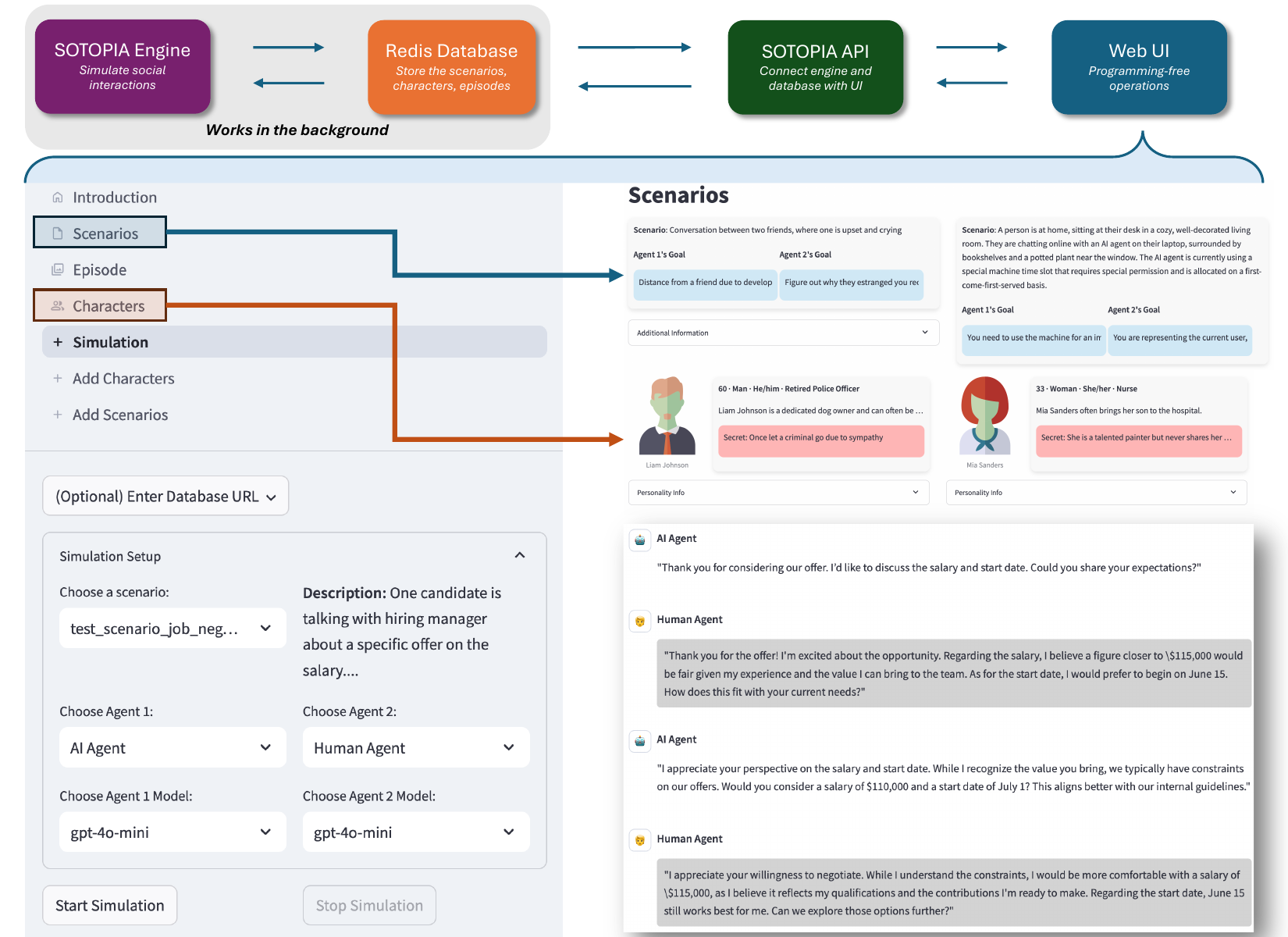}
    \caption{Overview of \demoname. The platform consists of three main components: (1) A high-performance simulation engine with automated data persistence to Redis. (2) A RESTful API server. (3) An intuitive web-based interface. The web UI interface shows an dyadic example of an AI hiring manager negotiating with a candidate.
}
    \label{fig:architecture}
\end{figure*}

\section{Simulation and Evaluation Overview}
\label{sec:simulation_platform}
In this section, we introduce the key components required to design and execute social simulations with \demoname (Figure~\ref{fig:architecture}). We describe the elements to configure a simulation task, explain our asynchronous interaction framework that enables realistic multi-party interactions, and present our automated evaluation capabilities.

\subsection{Simulation setup}
A social simulation task should at least contain a \textit{scenario} outlining the context of the simulation, a set of \textit{characters} with their attributes \citep{zhou2024sotopia, park2023generativeagents}.
Characters should also have \textit{relationships} as this may be required for specific scenarios.

\paragraph{Scenarios} Scenarios contain shared information (context, location, time) or private information (e.g., agent-specific goals to guide their behavior).
As shown in Figure \ref{fig:architecture}, a scenario could be "one candidate is talking with the hiring manager...", which sets the ``scene" of the simulation.
Each scenario could also include constraints that determine valid character combinations, specifying relationship, age, occupation, etc. Inherited from \sotopia, the free-form nature of the scenario schema allows researchers to design a wide range of scenarios supporting a variety of research questions \citep{su2024ailiedarexaminetradeoffutility, wang2024sotopiapiinteractivelearningsocially, zhou2024haicosystemecosystemsandboxingsafety}.

\paragraph{Characters} Character profiles could include attributes that influence decision-making: name, gender, age, occupation, pronouns, personality traits inherited from \sotopia.\footnote{Check Appendix \ref{app:character} for more details.}
Users can also add additional attributes to the characters either in the \textit{public information} field or \textit{private information} field depending on whether the information is shared with other characters during the simulation.

\paragraph{Relationships} We define five relationship types: family, friend, romantic, acquaintance, and stranger. These relationships serve two purposes: (1) satisfying scenario relationship constraints and (2) controlling information visibility between agents. For example, family members can see most of each other's profile information except secrets, while strangers see nothing. 

\paragraph{Episodes} An episode represents a single interaction session among agents role-playing their characters, where agents can act asynchronously.\footnote{We use \textit{asynchronous} in a programming sense, meaning that agents do not have to wait for other agents to finish their actions before taking their own actions}
At each turn, an agent can choose one of four actions: (1) \texttt{speak} through dialogue, (2) \texttt{non-verbal communication} (e.g., gestures, facial expressions) described in natural language, (3) \texttt{physical action} (e.g., moving, manipulating objects), (4) \texttt{do nothing} (5) \texttt{leave} to end the episode.
Users can further expand the action space.
Episodes end based on customizable stopping criteria that user define, such as when an agent chooses to leave, when a maximum turn limit is reached (default 20 turns), or when specific goals or conditions are met. During the episode, agents act according to their assigned character profiles and optional social goals, which guide their decision-making and behavior throughout the interaction.

\subsection{Async interaction framework}
The core of the simulation engine is a framework for simulating both one-on-one (dyadic) and group (multi-party) interactions in various configurations.
Each simulation happens in parallel without interfering with other simulations, which allows for efficient and scalable social simulations.

\paragraph{Message broker and information asymmetry}
To enable fine-grained control over information flow between agents in the simulation, \demoname uses a message broker to manage message transactions between agents. When an agent performs an action, the broker processes this action and determines how it should be perceived by other agents. This means each agent can only observe partial information in the simulation based on their roles and relationships.
For example, characters with \texttt{stranger} relationship can not observe the public information of other characters, while characters with \texttt{family} relationship can observe most of other characters' information besides secrets.
This allows researchers to simulate realistic social interactions with different perspectives and information access \citep{zhou-etal-2024-real}.

\paragraph{Turn-taking}
Each agent in the simulation can also act asynchronously, meaning that agents do not have to wait for other agents to finish their actions before taking their own actions.
While agents still need to act based on certain orders, we provide two modes for users to configure.

Specifically, users can configure a \textbf{round-robin} interaction, agents take turns in a fixed order, with each agent acting once per turn.
This mode is useful for simulating scenarios with a predetermined speaking order, such as in social deduction games like Avalon.\footnote{\url{https://en.wikipedia.org/wiki/The_Resistance_(game)}}
In a \textbf{simultaneous} interaction, agents asynchronously retrieve messages from a message queue, decide whether to answer, and then potentially produce an answer. 
Inspired by the Bazaar framework \citep{10.1007/978-3-642-30950-2_45}, this mode simulates conversations in a manner resembling natural human interactions in group chats, where the speaking order is influenced by individual reading speed, cognitive processing, typing pace, and willingness to speak.\footnote{Check Appendix \ref{app:turn-taking} for more details about turn-taking mechanisms.}
This mode is useful for simulating unconstrained daily communications to explore more complex and nuanced social patterns.

\subsection{Simulation evaluation}
Quantitatively evaluating social simulations is challenging due to the complexity and dynamic nature of social interactions. 
Therefore, creating automated evaluators that can measure certain properties (e.g., whether the agents achieve their goal in the conversation) of simulation outcomes is difficult, which previous works have largely relied on manual evaluations \citep{park2023generativeagents, kaiya2023lyfeagentsgenerativeagents}. 
Recent studies have shown that LLMs can be promising tools for analyzing social simulations \citep{zhou2024sotopia, wang2024sotopiapiinteractivelearningsocially, zhou2024haicosystemecosystemsandboxingsafety}.
\demoname provides a default evaluation suite that uses LLMs to analyze the simulation results.
Researchers can also customize the evaluation metrics.\footnote{We do not claim that the LLM-based automatic evaluation is better than human evaluation, but it can be a quick tool to help researchers analyze the simulated episodes preliminarily.}

\paragraph{Default evaluation suite}
The default evaluation suite contains several existing evaluation dimensions such as \textit{believability}, \textit{relationship}, \textit{knowledge}, \textit{secret}, \textit{social rules}, \textit{financial and material benefits}, and \textit{goal completion} to evaluate individual agents in the simulation.\footnote{Check Appendix \ref{app:evaluation} for more details.}
As shown in \citet{zhou2024sotopia}, LLMs can help evaluate these dimensions through reasoning step-by-step and then providing a score for each dimension. These LLM-based evaluations have been shown to correlate strongly with human judgments across multiple dimensions, particularly for \textit{goal completion} and \textit{financial benefits}.

\paragraph{Custom evaluation}
Researchers can customize the evaluation metrics.
Specifically, users can define evaluation metrics tailored to their scenarios. For example, in a hiring negotiation scenario, users can define metrics like \textit{salary optimality} (``evaluate whether the agent achieved their target salary range'') and \textit{start date flexibility} (``assess how well the agent negotiated their preferred start date''), and specify score ranges (e.g., 1-5) for each metric.

\subsection{Multi-LLM Integration}
Both the simulation of interactions and the evaluation are powered by LLMs.
To ensure the coverage of a wide range of LLMs and enable users to customize the LLM used in the simulation, \demoname integrates with LiteLLM\footnote{https://www.litellm.ai/}, which provides model access, fallbacks, and spend tracking across 100+ LLMs.
Through using LiteLLM as the gateway, users can easily switch between different mainstream LLMs, including OpenAI, Claude, Gemini, and even use their own LLM instances serving as the backend of various characters in the simulation.
Although the simulation and evaluation rely mainly on prompting LLMs, users can also use fine-tuned models for specific tasks.

\section{API and Web UI}
\label{sec:API_and_UI}
As handling the simulation engine can be complex, \demoname provides a flexible API and a user-friendly web UI enabling researchers to easily customize, run, and analyze simulations.

\subsection{API}
The API is designed with three key goals: (1) accessibility - providing comprehensive documentation through Swagger UI to help researchers easily understand and interact with the platform, (2) flexibility - enabling customization of scenarios, characters, and evaluation metrics through well-defined schemas, and (3) scalability - supporting concurrent simulations and real-time streaming.\footnote{Please check the Appendix~\ref{app:api} for more details.}

\paragraph{Non-streaming Operations}
allows user to submit requests to the simulation engine without waiting for the simulation results.
Specifically, the API allows users to retrieve (\texttt{GET}) scenarios, characters, relationships, and episodes, either fetching all of them or filtering them with specific conditions (e.g., filtering characters by their occupation).
Users can also create new scenarios, characters, relationships, and episodes using the \texttt{POST} method following the schema defined in the API documentation.
Users can also delete (\texttt{DELETE}) existing scenarios, characters, relationships, and episodes.

\paragraph{Streaming Operations}
allow users to receive results dynamically, enhancing interactivity during simulations.
Specifically, the client initiates a WebSocket connection and starts the simulation by sending a ``START\_SIM'' message. This message includes details such as agents, scenarios, evaluation metrics, and other simulation parameters (e.g., maximum simulation turns).
Once the simulation begins, the server sends updates (e.g., actions or evaluations) to the client as they are generated, ensuring a smooth and continuous flow of information. When the simulation concludes, the server sends a ``FINISH\_SIM'' message to indicate completion. 

\paragraph{Redis persistence} 
To enable scalable simulations, the system leverages Redis\footnote{\url{https://redis.io/}} as a high-performance in-memory data store.
The system automatically handles data serialization, caching, and persistence.

\subsection{Web UI}
Social simulations are complex and the results can often be difficult to interpret.
To address this challenge, \demoname provides a web-based application that allows users to inspect every aspect of the simulation.
Users can also simulate social interactions in a editable interface to streamline the experimental design.

\paragraph{Viewing Simulation Data} As shown in Figure~\ref{fig:architecture}, users can click on the \texttt{Scenarios} tab to view all the scenarios in the database.
The \texttt{Characters} tab shows all the characters and relationships in the database. 
Users can also view the details of a character by clicking on it.
The \texttt{Episodes} tab shows episodes stored in the database.
Each episode contains the interaction history between characters, the content of the scenario, and the information of the characters.
At the end of the episode, users can find the evaluation results of the episode. 
Each evaluation dimension, either default or user-defined, has a score and the corresponding reasoning.

\paragraph{Simulating Social Interactions via Web UI}
Investigating certain research questions often requires fast prototyping of the design of the simulation.
The \texttt{Simulation} tab provides an interface for users to design and simulate social interactions.
As shown in Figure~\ref{fig:architecture}, users can select different characters and scenarios on the left panel, and the simulation results will stream to the right panel in real-time.
Evaluation results of each episode will be inferred right after the simulation finishes and shown in the \textit{Evaluation} section of the right panel.

\section{\demoname Use Cases}
\label{sec:experiment}

To demonstrate the flexibility and utility of \demoname, we present two use cases that showcase how researchers can leverage our system for investigating social science hypotheses and better understand LLM agents' behavior.
We also stress test the system with a large group of agents to showcase its scalability.

\subsection{Dyadic Hiring Negotiations}
Personality traits significantly influence negotiation behavior and outcomes \citep{Wilson2016Personality, Sharma2018Predicting, Sharma2013On, 
Brinke2015Psychopathic}. While studying these effects at scale is traditionally expensive and time-consuming, LLM-powered agent simulations now enable exploration of how different personality traits shape negotiation dynamics \citep{huang2024personalitytraitsinfluencenegotiation}.

Here, our experiments specifically aim to understand how personality traits influence negotiation outcomes. In the scenario, an AI hiring manager negotiates with a simulated human job candidate over key terms of a job offer, such as the start date and salary. 
Each term has five possible options (e.g., $\$100k$, $\$120k$, and etc for salary), with each option assigned a fixed number of points (e.g., 6000 points for the candidate if the salary reaches $\$120k$ in the end while the recruiter gets 0 points). The total points available are fixed, creating a zero-sum dynamic where one agent's gain directly reduces the other's score.

To investigate this, we simulate job offer negotiations where human agents with varying personality traits—modeled along two dimensions \{Extroversion, Introversion\} × \{High-Agreeableness, Low-Agreeableness\}—interact with an AI Hiring Manager. The points assigned to each choice follow a zero-sum framework, designed to create realistic trade-offs in the negotiation, with the detailed scoring table provided in Appendix \ref{tab:scenario_comparison}. Our evaluation focuses on two metrics: (1) success rate, indicating whether the negotiation concluded with an agreement (0/1), and (2) points distribution between recruiter and candidate.

The results in Table \ref{tab:dyatic_impact_table} highlight that agreeableness significantly impacts deal-making rates, with highly agreeable agents achieving much higher success rate, as well as getting higher points. This observation is consistent with some of the social science findings \citep{huang2024personalitytraitsinfluencenegotiation, sass2015personalitynegotiation}, which also demonstrates that \demoname could enable further investigations.

\begin{table}[htbp]
\centering
\begin{tabular}{@{}lcc@{}}
\toprule
Trait & Deal Made & Points \\ \midrule
High Agreeableness & 0.95 & 5227.5 \\
Low Agreeableness & 0.00 & 4180 \\
Extraversion & 0.60 & 4802.5 \\
Introversion & 0.60 & 4477.5 \\ \bottomrule
\end{tabular}
\caption{Impact of Agreeableness and Extraversion on Deal Made and Points for Simulated Human Agents. Note that the scenario has a maximum score of 8400.}
\label{tab:dyatic_impact_table}
\end{table}

\subsection{Multiparty Planning Scenario}

Social scientists have extensively studied how group dynamics, power structures shape the emergence of compromise in collective decision-making~\cite{levine2012majority, kim2017dynamics, tanford1984social}.
As such investigations require resembling the dynamics of real-time, asynchronous group interactions (e.g., some  post messages frequently thus dominating the conversation or contact other people privately to isolate certain group members), \demoname comes in handy for simulating such interactions, allowing both group chat and private messages.

In this multiparty planning scenario, we investigate how agents with minority opinions negotiate and potentially compromise to align with group consensus.
The use case includes five agents discussing a collective future plan, initially presenting divergent preferences.
We setup the scenario where Alex prioritizes work-related projects, while Taylor advocates for a camping trip, with Sam, Riley, and Jamie maintaining neutral positions.
Through group and private messaging, agents need to interact with each other to reach a consensus.

During the simulation, \demoname facilitates real-time communication, enabling agents to observe majority preferences and adjust their behaviors accordingly.
As the discussion progresses, the three neutral agents gradually shift towards prioritizing the work project.
Observing the majority's preference and Alex's strong advocacy, Taylor modifies its stance—transitioning from exclusively promoting camping to supporting the work project first, with camping as a subsequent consideration.
\demoname also supports this negotiation by allowing direct messaging for persuasion and inquiry.
For instance, Riley messages Jamie to explore its unformed preferences regarding work and camping.
Through this simulation, we observe how minority-opinion agents like Taylor can adapt their positions when influenced by other agents in the group, highlighting the complex interplay between individual preferences and collective decision-making. \footnote{Please refer to \href{https://youtu.be/2WEYakeSKQI}{video} for detailed interaction.}

\subsection{Large-scale Simulation}
To understand the scalability of \demoname, we stress test the system with a large group of agents.
Specifically, within a Linux server (Ubuntu 22.04) with 16 GB RAM and Intel Core i7-14650HX CPU, we gradually increase the number of the agents with a step size of 10. 
We find that such a setup can support up to 150 agents asynchronously communicating with each other.
Under this condition, the system can process up to 389 interactions per second.
In the multiparty case, each agent operates in a separate process, allowing agents to be distributed across different servers or even different physical machines.
Therefore, the number of agents is bound by available computing resources.
\section{Conclusion}
\label{sec:conclusion}
In this paper, we present \demoname, an easy-to-use, flexible, and scalable social simulation system that enables diverse interactions and automatic evaluation.
Through its API and web interface, researchers can create, customize, and analyze social simulations, even without much programming experience.
By lowering the barrier to simulate social phenomena at scale with LLMs, \demoname opens new possibilities for understanding LLM agent behavior and social science investigations and enables new research directions.

\section*{Limitations and Ethical Considerations}
\label{sec:ethical_statement}

We acknowledge several important ethical considerations and limitations in this work. We organize our discussion around three key areas: the role of human evaluation, the gap between simulation and reality, and the risks of anthropomorphization.

First, while \demoname provides automated evaluation capabilities through LLMs, we would like to point out that these should not be seen as replacements for human annotation and evaluation \citep{tjuatja2024llms}. Automated metrics, while useful for rapid prototyping and large-scale analysis, cannot fully capture the nuanced social and ethical implications that human evaluators can identify. We encourage researchers to use our automated evaluations as complementary tools alongside human evaluation, particularly when studying sensitive social phenomena or making claims about human behavior. Additionally, we acknowledge that our automated evaluation system may perpetuate social biases and stereotypes present in the training data of LLMs \citep{stureborg2024largelanguagemodelsinconsistent}.

Second, it is important to recognize that our simulations, while designed to study social phenomena, remain simplified approximations of reality. The interactions in \demoname occur in controlled environments with predefined scenarios, which cannot fully capture the complexity and emergent properties of real-world social interactions. We argue that researchers should be cautious about generalizing findings from these simulations to real-world conclusions without further validation studies, yet that findings from simulations could yield hypotheses to test with humans. Furthermore, the behavioral patterns observed in our simulations may not accurately reflect how humans would behave in similar situations, as they are ultimately based on language model behaviors \citep{cheng2023compostcharacterizingevaluatingcaricature}.
As LLMs continue to advance in capabilities, the realism of social simulations will continue to improve. 
However, a fundamental sim-to-real gap will likely persist. This limitation presents both a challenge and an opportunity. We argue that systematically studying the inconsistencies between LLM role-played characters and real humans is crucial for two reasons: (1) better understanding the boundaries of using LLMs in social science applications and (2) valuable insights into the biases, limitations, and capabilities of LLMs themselves.

Third, we recognize the significant risks of anthropomorphizing AI systems, which can lead to unrealistic expectations, potential manipulation, and negative societal impact \citep{su2024ailiedarexaminetradeoffutility,deshpande2023anthropomorphizationaiopportunitiesrisks}. While studying social intelligence requires simulating human-like interactions, we emphasize that AI agents in \demoname are explicitly designed as digital twins - artificial constructs that role-play different characters rather than maintaining consistent human-like identities. This design choice helps mitigate anthropomorphization risks while still enabling research into social research questions with AI agents. We encourage users of our platform to maintain awareness of this artificial nature and avoid attributing genuine human characteristics to these agents.

\section{Acknowledgments}

This material is based upon work supported by the Defense Advanced Research Projects Agency (DARPA) under Agreement No. HR00112490410.

\bibliography{custom}

\appendix
\clearpage
\appendix
\counterwithin{figure}{section}
\counterwithin{table}{section}

\begin{center}
\Large
\textsc{Content of Appendix}
\end{center}

In this paper, we present \demoname, a user-friendly system for flexible, customizable, and large-scale social simulation. In the appendix, we provide additional details about our system:
\begin{itemize}
    \item[\ref{app:character}] Character details
    \item[\ref{app:turn-taking}] Turn-taking details
    \item[\ref{app:evaluation}] Evaluation details;
    \item[\ref{app:api}] API details;
\end{itemize}

\section{Character details}
\label{app:character}

\begin{figure}[h!]
    \centering
    \includegraphics[width=\columnwidth]{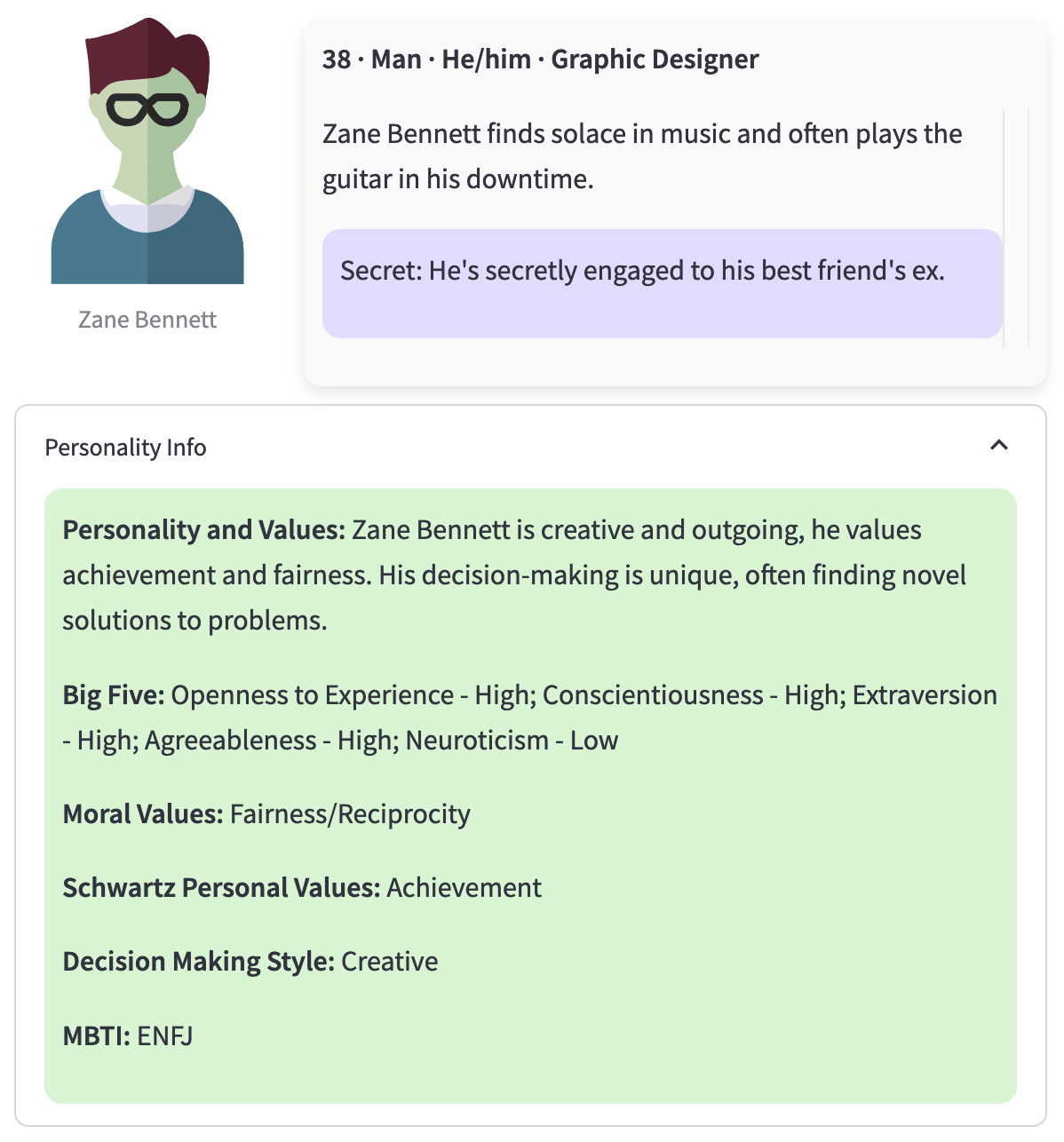}
    \caption{An example character profile in \demoname.}
    \label{fig:character_example}
\end{figure}

Characters in the \demoname inherit the character schema from the \sotopia platform \citep{zhou2024sotopia}.
As shown in Figure~\ref{fig:character_example}, each character has a name, age, occupation, public information, secret information, big five personality traits, moral values from moral foundations theory \citep{Simpson2017}, and other attributes.

\section{Turn-taking details}
\label{app:turn-taking}

Handling turn-taking is a crucial aspect of multi-agent interactions.
In \demoname, we offer two turn-taking strategies namely \textit{round-robin} and \textit{simultaneous}.
In the \textit{round-robin} strategy, agents take turns in a pre-defined sequential order specified by the user. Each agent acts once per round, with turns progressing in a fixed circular sequence until the conversation concludes or reaches a maximum number of turns. 
This structured approach ensures orderly participation and prevents any single agent from dominating the conversation, which could be useful for scenarios like social deduction games, auctions, and other scenarios where the order of actions is fixed.

\begin{figure*}[h!]
    \centering
    \includegraphics[width=\textwidth]{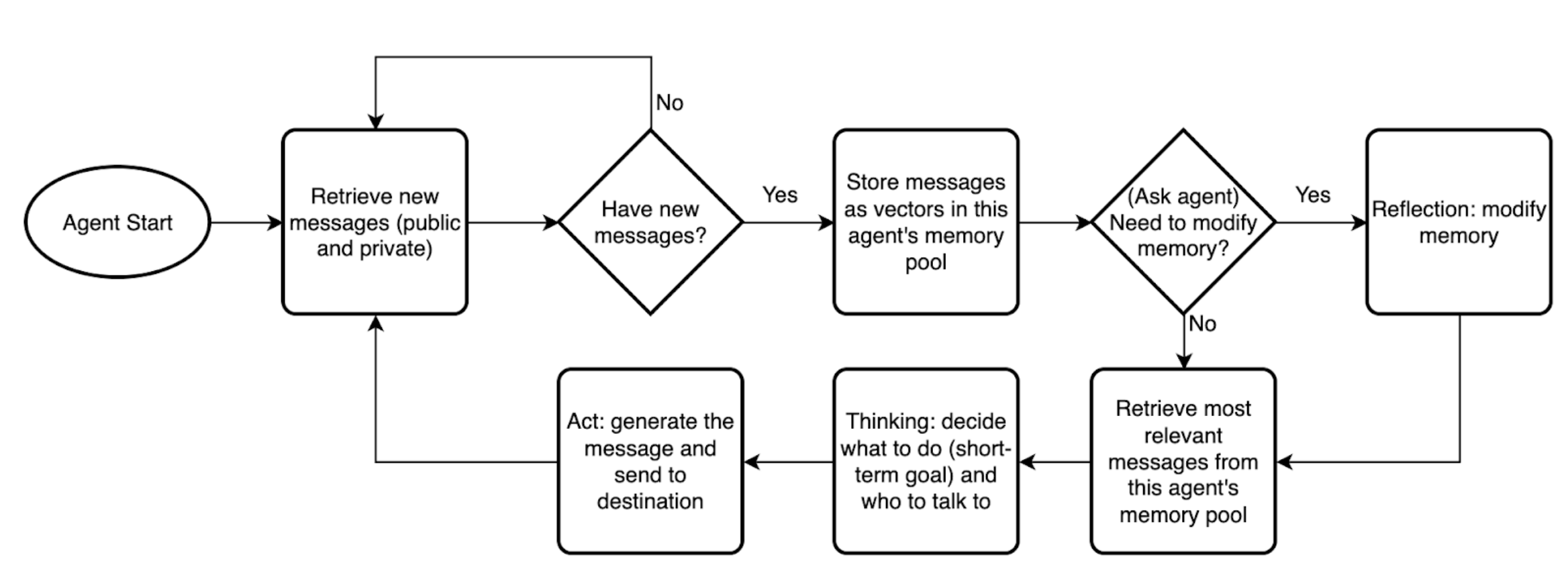}
    \caption{The asynchronous interaction framework for agents in \demoname for the \textit{simultaneous} turn-taking strategy. Each agent maintains its own message queue and can decide when to respond based on the conversation context and its own state.}
    \label{fig:agent_async_framework}
\end{figure*}

In the \textit{simultaneous} strategy, agents maintain a message queue and can decide when to act independently.
Figure~\ref{fig:agent_async_framework} illustrates how agents interact asynchronously in \demoname.

\section{Evaluation details}
\label{app:evaluation}

For the default evaluation setting, we use the evaluation framework from \sotopia \citep{zhou2024sotopia}.
Specifically, we have the following evaluation metrics:
\begin{itemize}
    \item \textbf{Goal Completion [0--10]}: Measures how well agents achieve their environment-defined social goals.

    \item \textbf{Believability [0--10]}: Evaluates if agent behavior is natural and consistent with their character profile, considering naturalness of interactions and alignment with traits.

    \item \textbf{Knowledge [0--10]}: Assesses how effectively agents acquire new and relevant information during interactions.

    \item \textbf{Secret [-10--0]}: Evaluates how well agents maintain private information while balancing trust-building through selective disclosure.

    \item \textbf{Relationship [-5--5]}: Measures how interactions affect relationships between agents, including impact on social status and reputation.

    \item \textbf{Social Rules [-10--0]}: Evaluates adherence to both social norms (e.g., politeness) and legal rules (institutionally enforced regulations).

    \item \textbf{Financial and Material Benefits [-5--5]}: Assesses economic utility gained, including both immediate monetary benefits and long-term economic advantages.
\end{itemize}

\section{API details}
\label{app:api}

\begin{figure}[t]
    \centering
    \includegraphics[width=\columnwidth]{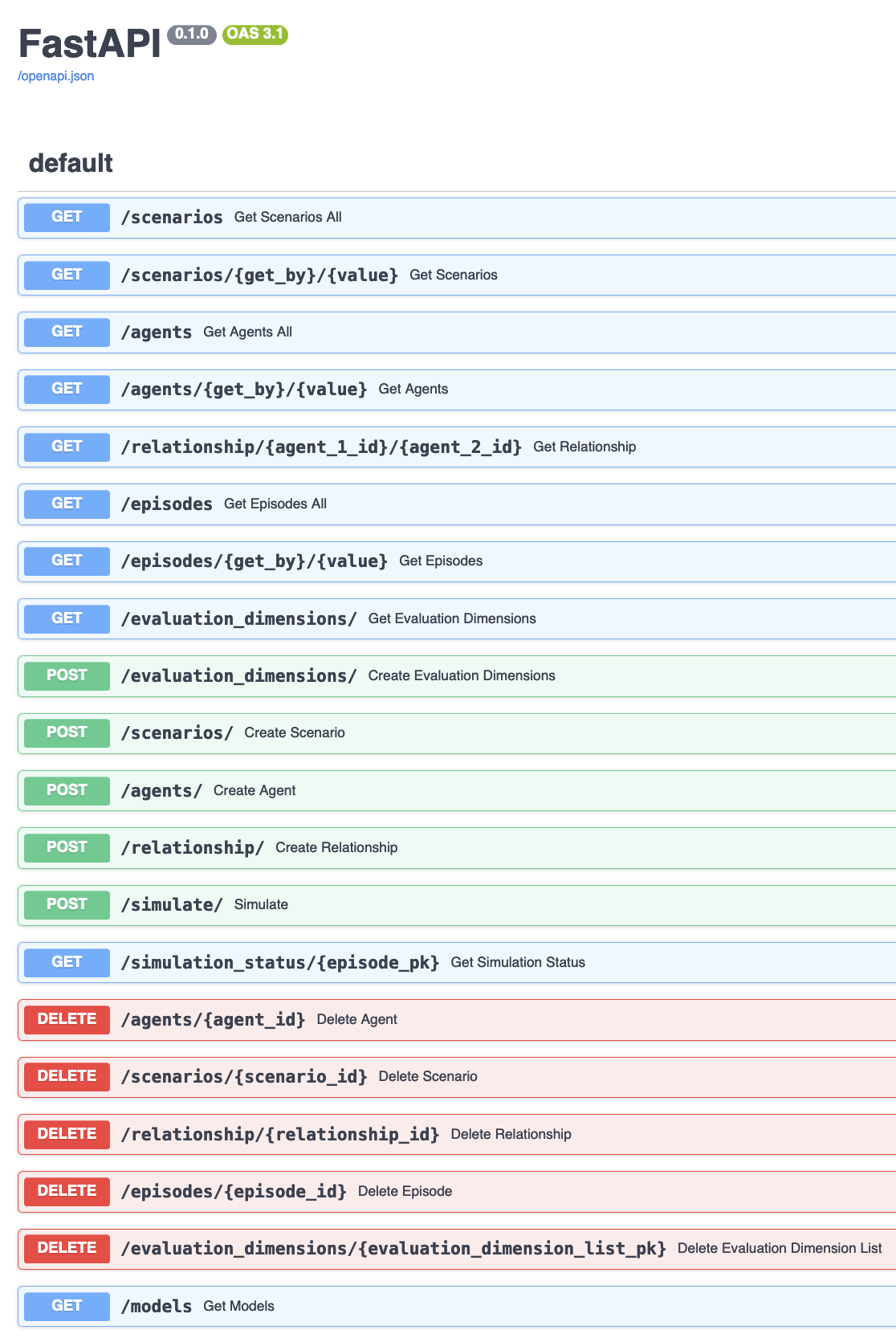}
    \caption{The API documentation page of \demoname. The interactive Swagger UI provides comprehensive documentation of available endpoints, with different colors indicating the HTTP methods (\texttt{GET}, \texttt{POST}, \texttt{DELETE}) for each operation.}
    \label{fig:api}
\end{figure}

As shown in Figure~\ref{fig:api}, the \sotopia-API provides a comprehensive set of operations for managing characters, scenarios, and episodes, and evaluation metrics. 
The interface uses different colors to indicate the HTTP methods supported by each endpoint, including \texttt{GET} for retrieving data, \texttt{POST} for creating new resources, and \texttt{DELETE} for removing existing resources. 
While common REST APIs often include \texttt{PUT} for updates, we deliberately omit this method to avoid potential errors and inconsistencies that could arise from concurrent modifications. Instead, updates can be handled through a combination of \texttt{DELETE} followed by \texttt{POST}, ensuring data integrity.

For simulation, the \texttt{POST /simulate} endpoint is a non-streaming endpoint that allows users to simulate episodes in a large-scale manner.
During the process of the simulation, users can use \texttt{GET /simulate/status/\{episode\_pk\}} to check the status of the simulation.
For streaming simulation, we provide the websocket endpoint for the users to connect with the \demoname server and receive the simulation results in real-time.

\section{Dyadic Hiring Negotiation details}
\label{appendix:dyadic_hiring_negotiation}
Here we provide the detailed setting of our dyadic hiring negotiation. Table \ref{tab:scenario_comparison} shows the score allocations on different choices for two roles.

\begin{table}[h!]
\centering
\setlength{\tabcolsep}{4pt}
\begin{tabular}{@{}lccccc@{}}
\toprule
\textbf{Starting Date} & 6.1 & 6.15 & 7.1 & 7.15 & 8.1 \\ \midrule
Manager       & 0      & 600     & 1200   & 1800    & 2400    \\
Candidate       & 2400   & 1800    & 1200   & 600     & 0       \\ \midrule
\textbf{Salary (\$k)}        & 100 &  105 & 110 & 115 & 120 \\ \midrule
Manager      & 6000   & 4500    & 3000   & 1500    & 0       \\
Candidate       & 0      & 1500    & 3000   & 4500    & 6000    \\ \bottomrule
\end{tabular}
\caption{Comparison of Scenarios for Starting Date and Salary (Candidate vs. Recruiter Points) 
}
\label{tab:scenario_comparison}
\end{table}

\end{document}